\documentclass[%
reprint,
superscriptaddress,
 amsmath,amssymb,
 aps,
]{revtex4-2}

\usepackage{graphicx}
\usepackage{dcolumn}
\usepackage{bm}
\usepackage{xcolor}
\graphicspath{{Figures/}}
\usepackage{gensymb}

\begin{document}

\preprint{}

\title{Excitation of the lower-hybrid drift instability in the outflow of electron-only magnetic reconnection}

\author{B. K. Russell}%
 \email{br2114@princeton.edu}
\affiliation{Department of Astrophysical Sciences, Princeton University, Princeton, New Jersey 08544, USA}

\author{K. Sakai}%
\affiliation{National Institute for Fusion Science, 322-6 Oroshicho, Toki, Gifu 509-5292, Japan}

\author{Y. Zhang}%
\affiliation{Department of Astrophysical Sciences, Princeton University, Princeton, New Jersey 08544, USA}
\affiliation{University Corporation for Atmospheric Research, Boulder, CO, 80301, USA 
}

\author{L. Gao}%
\affiliation{Princeton Plasma Physics Laboratory, Princeton University, 100 Stellarator Rd, Princeton, NJ, 08540, USA}

\author{E. G. Blackman}%
\affiliation{Department of Physics and Astronomy, University of Rochester, Rochester, NY 14627, USA}

\author{W. Daughton}%
\affiliation{Los Alamos National Laboratory, Los Alamos, New Mexico 87545, USA}

\author{C. Dong}%
\affiliation{Center for Space Physics and Department of Astronomy, Boston University, Boston, MA 02215, USA}
\affiliation{School of Natural Sciences, Institute for Advanced Study, Princeton, NJ 08540, USA}

\author{J. Katz}%
\affiliation{University of Rochester Laboratory for Laser Energetics, Rochester, NY, USA}

\author{S. R. Klein}%
\affiliation{Nuclear Engineering and Radiological Sciences, University of Michigan, Ann Arbor, Michigan 48109, USA}

\author{C. C. Kuranz}%
\affiliation{Nuclear Engineering and Radiological Sciences, University of Michigan, Ann Arbor, Michigan 48109, USA}

\author{X. Li}
\affiliation{Los Alamos National Laboratory, Los Alamos, New Mexico 87545, USA}

\author{X. M. Li}%
\affiliation{Center for Space Physics and Department of Astronomy, Boston University, Boston, MA 02215, USA}

\author{A. L. Milder}%
\affiliation{University of Rochester Laboratory for Laser Energetics, Rochester, NY, USA}

\author{J. Ng}
\affiliation{Department of Astronomy, University of Maryland, College Park, Maryland 20742, USA}
\affiliation{NASA Goddard Space Flight Center, Greenbelt, Maryland 20771, USA}

\author{K. Orr}%
\affiliation{Department of Mechanical and Aerospace Engineering, Princeton University, Princeton, New Jersey 08544, USA}

\author{G. Pomraning}%
\affiliation{Department of Astrophysical Sciences, Princeton University, Princeton, New Jersey 08544, USA}

\author{J. P. Schell}%
\affiliation{Nuclear Engineering and Radiological Sciences, University of Michigan, Ann Arbor, Michigan 48109, USA}

\author{A. Stanier}
\affiliation{Los Alamos National Laboratory, Los Alamos, New Mexico 87545, USA}

\author{J. Yoo}%
\affiliation{Princeton Plasma Physics Laboratory, Princeton University, 100 Stellarator Rd, Princeton, NJ, 08540, USA}

\author{H. Ji}%
\affiliation{Department of Astrophysical Sciences, Princeton University, Princeton, New Jersey 08544, USA}
\affiliation{Princeton Plasma Physics Laboratory, Princeton University, 100 Stellarator Rd, Princeton, NJ, 08540, USA}

\date{\today}

\begin{abstract}

We report experimental evidence for the lower-hybrid drift instability in the current sheet normal direction of electron-only magnetic reconnection. In our laser-driven capacitor-coil experiment, the system size ($\sim$3 ion skin depths) places it in the electron-only regime. Yet, Thomson scattering reveals out-of-plane electron drift oscillations at the local lower-hybrid frequency, with kinetic energy density reaching $\sim$18\% of the local magnetic energy density.  Linear theory with the measured parameters predicts more than ten e-folding times of growth, indicating that the instability reaches the nonlinear regime within the measurement window. Supported by particle-in-cell simulations, these results demonstrate the importance of ions in the dissipation and energy transfer in electron-only reconnection where their significance has not been previously recognized.

 \end{abstract}

\maketitle

As a fundamental process in the universe, magnetic reconnection has been studied extensively, including in-situ satellite observations \cite{burch16,Phan_Nature_2018_e_only_magnetosheath}, large-scale simulations \cite{Guo_2015,Dong_SA_2022}, and dedicated laboratory experiments \cite{Nilson_PRL_2006,Hare_PRL_2017,ji23}. In this process, changes in magnetic field topology  allow for the rapid conversion of plasma magnetic to kinetic energy. This can power energetic phenomena in sufficiently magnetized astrophysical sources and drive disruptive events in laboratory fusion plasmas~\cite{Ji_POP_2011}. The reconnection current sheet and outflow are complex plasma environments with various sources of free energy in the form of plasma flows, density and field gradients, and strong electric currents. As a result, a large set of instabilities can occur and various waves can be generated throughout the reconnection region. These waves can dissipate currents, act as a source of anomalous dissipation and diffusion, and modify particle distributions. Several of these waves have been measured in-situ by the Magnetospheric Multiscale (MMS) mission as recently reviewed by Graham \textit{et al.} \cite{graham_SSR_2025}. 

Of the possible waves that can be generated, lower-hybrid drift waves (LHDW) have been found to develop in most regions with varying impact on the reconnection dynamics and energy dissipation. Near the current sheet, electron drifts perpendicular to the in-plane magnetic field can generate LHDW in the reconnection out-of-plane direction through the lower-hybrid drift instability (LHDI). Recent measurements of the Earth's magnetotail \cite{Chen_PRL_2020} demonstrate that LHDW can drive vortical flows in the electron diffusion region (EDR) of electron-ion reconnection. Laboratory experiments \cite{Yoo_PRL_2024,Yoo_PoP_2025} on the Magnetic Reconnection Experiment (MRX) demonstrated that electrostatic LHDW in the EDR account for approximately 20\% of the reconnection electric field. 

\begin{figure}
\centering
\includegraphics{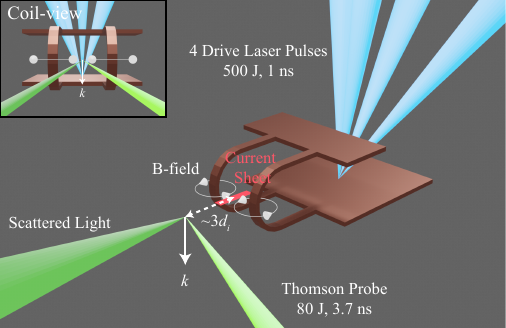}
\caption{Experimental setup for Thomson scattering on OMEGA to diagnose the reconnection out-of-plane waves. A Cu capacitor coil target is driven by 4 laser pulses generating a reconnection geometry. The outflowing plasma from the reconnection is diagnosed by Thomson scattering.}
\label{Fig:setup}
\end{figure}

While LHDW have shown to be important in electron-ion reconnection, even near the EDR \cite{Ahmadi_JGR_2025}, their importance in electron-only reconnection remains unclear with limited observational evidence from MMS showing oscillations in the electric and magnetic field near the lower-hybrid frequency \cite{Wang_JGR_2022}. In this form of reconnection, an EDR is formed with an electron outflow. However, due to spatial and/or temporal constraints, there is no associated ion outflow. While originally measured in the Earth's turbulent magnetosheath \cite{Phan_Nature_2018_e_only_magnetosheath}, this form of reconnection has since been measured in many systems \cite{Man_GRL_2020,Liu_JGR_2020,Hubbert_JGR_2022}, and has been the focus of various theoretical \cite{Pyakurel_PoP_2019,Liu2025_eonly}, and some experimental studies \cite{Shi_PRL_2022,Greess_PoP_2022,Shi_PRL_2023,Chien_NatPhys_2023,Zhang_NatPhys_2023}. 

In this Letter, we show experimental evidence of LHDW in the out-of-plane direction of the outflow from electron-only reconnection. The system size is approximately three ion skin-depths for the measured Cu ion plasma, placing it firmly in the electron-only regime \cite{Pyakurel_PoP_2019}. Here, the LHDW are generated by strong cross-field drifts formed primarily by reconnection outflows dominated by electron motion. Thomson scattering measurements show large amplitude swings in the out-of-plane electron velocity with frequency peaked at the local lower-hybrid frequency. Dispersion relation calculations with the measured plasma parameters demonstrate that the system is lower-hybrid drift unstable with the experimental current sustained for $>10$ e-folding times. This result directly demonstrates the importance of ions, even in electron-only reconnection where they have been believed to have a limited impact. For turbulent systems, an important implication is that electron-only reconnection, which can be a sink for energy from the turbulent cascade, may be able to re-inject energy near the scale of the electron Larmor radius.

\begin{figure*}
\centering
\includegraphics{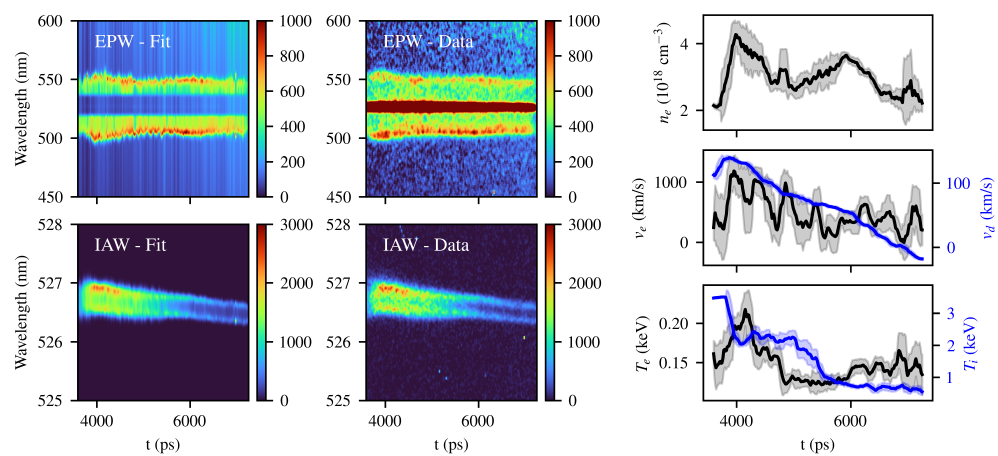}
\caption{Streaked Thomson scattering data and simultaneous EPW/IAW fits performed with Tsadar used to extract the parameters shown on the right side of the figure. Fits were performed with a super-gaussian with an index of 3.5 as expected for the probe conditions \cite{Matte_1988} and assuming a constant charge state Z = 20 to better constrain the simultaneous EPW/IAW fits. The extracted parameters are plotted as a rolling average (solid lines) over 150 ps and a standard deviation (filled region) over this same duration.}
\label{Fig:TS}
\end{figure*}


The experimental data presented here was obtained on the OMEGA laser system at the University of Rochester Laboratory for Laser Energetics. Similar to our previous experiment \cite{Zhang_NatPhys_2023}, a capacitor coil target created by bending a 50 $\mu$m thick laser cut Cu foil was irradiated by several lasers (see Fig. \ref{Fig:setup}). While six beams were used previously, here four 500 J, 1 ns duration, 351 nm central wavelength, pulses irradiated the backplate. The laser pulses heat the backplate, generating a plasma plume and an electrostatic potential that pulls electrons from the front to the backplate. This return current produces azimuthal magnetic fields around the legs of the coil target---forming a magnetic reconnection geometry. These fields and the reconnection dynamics have previously been diagnosed using proton radiography showing the formation of $\sim 50$ T magnetic fields in the reconnection upstream \cite{Chien_POP_2019}. Plasma flows into the region between the coils from both the backplate and the coil legs that are heated by the x-rays from the laser interaction. A secondary laser scatters from the plasma in the region between the coil legs and $600\;\mu$m from the top of the coils. The probe beam wavelength (526.5 nm) is offset from the driving lasers, and appears for a duration of 3.7 ns, providing Thomson scattering measurements over a large temporal window. A 100 $\mu$m distributed phase plate (DPP) was used and the probe energy was reduced to 80 J, resulting in a larger signal-to-noise ratio than in our previous measurements \cite{Zhang_NatPhys_2023}. By changing the probe and driver timing we were able to capture the evolution of the plasma from $t_0+1$ ns to $t_0+12$ ns where $t_0$ is the time of laser incidence on the backplate. For the purpose of this Letter, we will focus on the dynamics from approximately 3.5 -- 7 ns where the scattering signal is sufficiently above the background and the plasma is rapidly evolving.

Compared to our previous experiment where the Thomson probe measured the waves generated $17 \degree$ from the reconnection outflow \cite{Zhang_NatPhys_2023}, here the measured waves are exactly in the reconnection out-of-plane for the current sheet that forms between the top of the coils. The probe scatters from this region and is collected at $63\degree$ into two spectrometers coupled to streak cameras with a large spectral window and narrow spectral window providing measurements of the electron plasma wave (EPW) and ion acoustic wave (IAW) scattered signal respectively. The scattered light is imaged with a magnification of 2.1 and a 100 $\mu$m pinhole is used, giving a scattering volume of approximately $50\times 50 \times90\;\mu$m$^3$. Additionally, the transmitted beam is imaged, providing a clear diagnostic of whether the beam is filamented or strongly refracted. In the following we detail the evolution of the Thomson scattering signal. 

The Thomson scattering raw and fitted data taken over the temporal window from approximately 3.5 to 7 ns are shown in Fig.~\ref{Fig:TS}. Fitting was performed using the code Tsadar which employs auto-differentiation and is compatible with GPUs, allowing it to rapidly fit spectra with several parameters \cite{Milder_2024}. Fits were performed simultaneously for the EPW and IAW approximately every 15 ps and were used to extract the plasma properties. This method allows us to capture oscillations in the parameters, however it is also susceptible to high frequency noise that cannot be physical due to the limited temporal resolution of the system. While the streaking coupled to the detector has an inherent blurring, the largest contribution to the temporal resolution comes from pulse front tilt that limits the resolution to approximately 150 ps \cite{Swadling_RSI_2022}. To obtain the temporal evolution shown in Fig.~\ref{Fig:TS} of the parameters we therefore perform a rolling average over 150 ps and capture the standard deviation of the parameters in this temporal window. 

\begin{figure}
\centering
\includegraphics{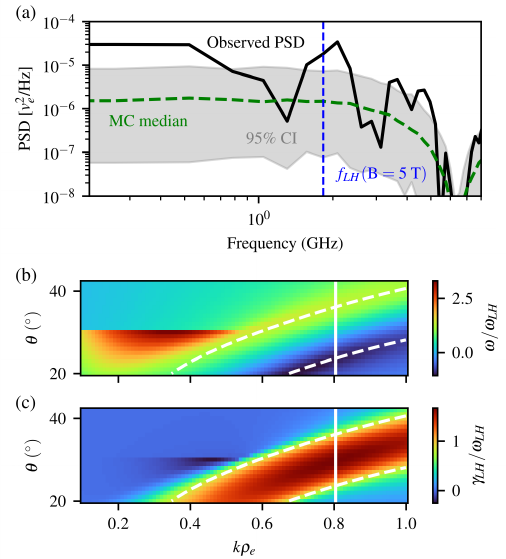}
\caption{Measured frequencies from the electron drift velocity. Periodogram (solid line), 95\% confidence interval (gray band), and lower hybrid frequency at 5 T (blue dashed line) are shown. Plot (b) shows the frequency and (c) the growth rate of the unstable modes generated by the LHDI as a function of the angle $\theta$ with respect to the electron propagation direction and the wave number $k$ multiplied by the electron Larmor radius $\rho_e$. Dashed white lines are plotted at half the maximum growth rate for reference. The solid white line shows the size of the Thomson scattering probe. }
\label{Fig:disp}
\end{figure}

The fitted Thomson scattering parameters demonstrate that the reconnection outflow evolves rapidly with strong oscillations in the electron drift velocity during this time. We observe the rapid growth of the electron density peaking at approximately $4\times10^{18}$ cm$^{-3}$. Coincident with this density peak is a large plasma flow velocity of $v_d>100$ km/s and an electron velocity $v_e>1000$ km/s. These velocities are both directed along the reconnection out-of-plane and the electrons have the same directionality that would be expected for those forming the current sheet. We note that while the electron temperature, density, and velocity are similar to those previously reported where unstable IAW were generated, here the ion temperature is measured earlier in time and is almost an order of magnitude higher. The large
drift velocity stabilizes the system to the ion-acoustic instability (see Appendix A for dispersion calculation). While we would therefore not expect to see signatures of the IAI, we do observe swings in the electron velocity approaching 1000 km/s on sub-nanosecond timescales. These swings are calculated from the evolution of the ratio between the Stokes and anti-Stokes peaks of the IAW data. While this measurement is susceptible to noise, our method to time average over several fits and plot the standard deviation accurately quantifies contributions from this noise. Even with the standard deviation from this noise, there is still a clear oscillation. This measurement may also be affected by stimulated Brillouin scattering \cite{Turnbull_PRL_2026}, however fits were performed with and without it included and it was found to be negligible for the probe parameters used here.

To quantify the oscillation frequency we calculate the periodogram shown in Fig.~\ref{Fig:disp}(a). To include the error we generate 500 sets of time-series data randomly sampled at each time from a normal distribution with the standard deviation taken from the $v_e$ curve and averaged over a 150 ps window. By similarly transforming this data we create a 95\% confidence interval band in frequency space, above which signal is unlikely to be resulting from correlated noise. There is a clear peak above this confidence interval that coincides with the lower-hybrid frequency $f_{LH}$ with a magnetic field of $\sim$5 T. While the magnetic field may be varying significantly in the probe region, especially with such strong currents, 5 T is similar to the strength calculated at the probe position based on the Biot-Savart law for the realistic 3D coil geometry in vacuum which gives a value of $\sim 6.5$ T.   

\begin{figure*}
\centering
\includegraphics{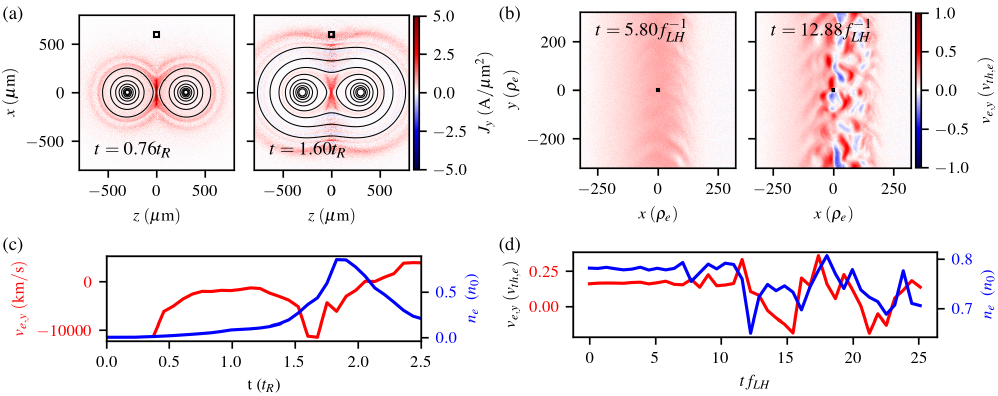}
\caption{2D VPIC simulations of (a) current driven reconnection and (b) the LHDI driven by current flowing perpendicular to magnetic fields. In (a) and (b) the black square is the Thomson probing position and contours of constant $A_y$ are shown in (a). Plots (c) and (d) demonstrate the electron velocity and density taking from the probe positions in (a) and (b) respectively.}
\label{Fig:sim}
\end{figure*}

Using parameters similar to those from the Thomson scattering analysis at the peak ($\sim 4$ ns, $v_e = 900$ km/s, $B = 5$ T, $n_e = 4\times10^{18}$ cm$^{-3}$, $T_i=2250$ eV, $T_e=180$ eV) we solve the 2D dispersion relation for lower-hybrid waves \cite{Yoo_PoP_2022}. The results of this dispersion calculation are shown in Fig.~\ref{Fig:disp}(b-c). The system is found to be highly unstable to the LHDI with a peak growth rate $\gamma_{LH} = 1.6\omega_{LH}$. The growth rate is large over a significant bandwidth of frequencies and wavenumbers with the peak frequency around $0.4\omega_{LH}$. The growth rate in this region is large with an e-folding time of approximately 0.05 ns. In Fig.~\ref{Fig:TS} the initial peak in $v_e$ lasts for approximately 0.5 ns or 10 e-folding times. We therefore expect that the instability enters the nonlinear regime giving the large swings in the drift velocity. Notably the growth rate is similar to that of the IAI previously reported in-plane at 7 ns \cite{Zhang_NatPhys_2023}. It is therefore expected that these instabilities should be growing on similar timescales, however in different directions. Importantly, the growth rate of the LHDI is large for wavelengths larger than the size of the Thomson scattering probe ($50\;\mu$m), denoted by a solid white line. Oscillations to the left of this line can be measured, while shorter wavelengths may be averaged over. Due to the large bandwidth of unstable frequencies, many frequencies are expected to exist simultaneously. However, the fastest growing modes are not aligned with the probe axis, therefore the measured frequencies and amplitude of oscillations in the electron velocity may be reduced from the true values. Furthermore, because the waves are propagating they could come from various regions of the outflow or even the edge of the current sheet where LHDI is more commonly found. The narrow bandwidth of the measured oscillations around $f_{LH}$ expected in the probe region suggests that these values are generated close to the probed volume. 

To understand the structure of the reconnecting plasma and the source of the currents in the region of the probe, 2D particle-in-cell simulations were run in the code VPIC \cite{Bowers_PoP_2008}. These simulations are based off of those of Chien \textit{et al.} \cite{Chien_NatPhys_2023}. Two regions of out-of-plane current are initialized with a linear ramp over a time $t_R$ followed by an exponential decay. An initial background plasma is included and plasma is injected around the coils during the ramp-up time. A reduced plasma frequency to gyrofrequency and increased electron and ion temperature is used to reduce the computational cost (see Appendix B for simulation details). During the ramp-up time, the injected plasma is driven strongly into the mid-plane where a current sheet is formed and reconnection occurs. From previous characterization of nanosecond laser-driven capacitor coils \cite{Chien_POP_2019}, the ramp-up time is approximately equal to the laser pulse duration. From the simulations we find that a significant outflow and out-of-plane electron velocity does not occur in the region of the probe until well into the decay time. This is qualitatively consistent with the experimental measurement which occurs approximately 3 ns after the end of the laser interaction. The simulation demonstrates that during this time push-type reconnection is still occurring despite the decay of the coil current. Furthermore, it predicts two sources of current within the measurement volume: a short-duration return current, and a more broadly distributed current. The cross-field electron drift that forms these currents is composed of both diamagnetic and $\mathbf{E}\times \mathbf{B}$ drifts. However, due to the scaling of the simulation the electric field strength and therefore the $\mathbf{E}\times \mathbf{B}$ drift is over-predicted.

To understand the evolution of the LHDI in this configuration we perform a 2D simulation in VPIC using the setup demonstrated by Ng \textit{et al.} \cite{Ng_PRR_2024}. The simulation is setup as a pressure balance with the density, magnetic field, and temperature varying along the x-direction. An electron drift is initialized in the middle of the box. The realistic electron drift velocity normalized to the thermal velocity is used. This drift velocity is set in the simulation by an electron density gradient; however, more generally in the experiment this cross-field drift should be driven by the diamagnetic and $\mathbf{E}\times \mathbf{B}$ drift. The plasma beta in the left side of the simulation space is also maintained. To reduce the computational cost we use ions with half the mass of a proton and increase the magnetic field strength, increasing the temperature to maintain $\beta$ while reducing the ratio of plasma frequency to gyrofrequency (see Appendix B for additional details). We therefore do not expect the growth rate and saturation of the instability to be correct. However, we can obtain a qualitative picture from this simulation while using the values from Fig.~\ref{Fig:disp} to understand the realistic growth rate and frequency.

In the simulation the instability grows linearly over a period of $\sim5.8f_{LH}^{-1}$ before becoming highly nonlinear. The linear period sees the formation of a well defined wave (Fig.~\ref{Fig:sim}(b) $t = 5.8f_{LH}^{-1}$) with an oscillating longitudinal electric field ($E_y$) and electron velocity. In the nonlinear period (Fig.~\ref{Fig:sim}(b) $t=12.88f_{LH}^{-1}$)  the electron velocity oscillates significantly with swings of 100\% (Fig.~\ref{Fig:sim}(d)). However, the density does not oscillate significantly with jumps of around 10\%. This qualitatively matches the experimental data in Fig.~\ref{Fig:TS} where the velocity starts approximately constant before oscillating significantly, while the density does not.

In conclusion, we have presented experimental evidence consistent with the growth of the LHDI in the outflow of electron-only reconnection. In our conditions, the ion Larmor radius, $\rho_i\approx390\; \mu$m, is comparable to the $600\;\mu$m separation of the coils, which is also $\sim 3$ ion skin depths, placing it in the electron-only regime. Despite this, the measured out-of-plane electron drift oscillates at the local lower-hybrid frequency with an amplitude approaching 100\% and a kinetic energy density $\sim18\%$ of the local magnetic energy density. Taken together with our previous results in the near-in-plane direction, where bursty IAW were measured \cite{Zhang_NatPhys_2023}, we find that ions may play a significant role in the reconnection outflow, despite their unimportance within the current sheet. Interestingly, LHDW have also been observed by MMS in electron-only reconnection \cite{Wang_JGR_2022}.

The local $\beta$ in our measurement region is much larger than that in the reconnection upstream and is comparable to the conditions in the turbulent magnetosheath in which electron-only reconnection is observed, so it is natural to ask whether the same coupling operates there. If it does, electron-only reconnection---usually regarded as a sink for energy---could couple to ions in the outflow and re-inject wave energy at scales near the electron Larmor radius, with wave-vectors nearly perpendicular to the local magnetic field. The smaller separation between current sheets in electron-only reconnection may also allow such waves to interact with neighboring layers and occupy a larger fraction of the volume than in standard electron-ion reconnection. Testing this will require measurements that resolve the wave-vector as well as the frequency, which is beyond the scope of the present work.

This work was supported by the US Department of Energy High-Energy-Density Laboratory Plasma Science program under Grant No. DE-SC0020103. The experiment was conducted at the Omega Laser Facility with the beam time through the National Laser Users' Facility user program. This material is based upon work supported by the Department of Energy [National Nuclear Security Administration] University of Rochester ``National Inertial Confinement Fusion Program'' under Award Number(s) DE-NA0004144. Work performed under the auspices of the U.S. Department of Energy by General Atomics under NNSA Contract 89233124CNA000365. Y.Z. was supported by the NASA Living with a Star Jack Eddy Postdoctoral Fellowship Program, administered by UCAR's Cooperative Programs for the Advancement of Earth System Science (CPAESS) under award $\#$80NSSC22M0097. C.D. was supported by DOE grant DE-SC0024639, NSF grant AGS-2301338, the Alfred P. Sloan Research Fellowship, and the IBM Einstein Fellow Fund at the Institute for Advanced Study, Princeton.

This report was prepared as an account of work sponsored by an agency of the United States Government. Neither the United States Government nor any agency thereof, nor any of their employees, makes any warranty, express or implied, or assumes any legal liability or responsibility for the accuracy, completeness, or usefulness of any information, apparatus, product, or process disclosed, or represents that its use would not infringe privately owned rights. Reference herein to any specific commercial product, process, or service by trade name, trademark, manufacturer, or otherwise does not necessarily constitute or imply its endorsement, recommendation, or favoring by the United States Government or any agency thereof. The views and opinions of authors expressed herein do not necessarily state or reflect those of the United States Government or any agency thereof.

The data that support the findings of this study are available from the corresponding author upon reasonable request.

\bibliography{compressed_bib}

\section*{End Matter}

\textit{Appendix A: Stability to IAW} -- While our previous experiment demonstrated the rapid growth of ion and electron acoustic waves here we demonstrate that for the parameters measured from approximately 3.5 to 7 ns in the out-of-plane direction the system is ion acoustic stable at all times. We solve the dispersion relation outlined previously \cite{Zhang_NatPhys_2023}, 
\begin{multline}
    1 - \frac{\omega_{pe}^2}{2k^2T_em_e}Z'\left(\frac{\omega/k-v_d}{\sqrt{2T_e/m_e}}\right) \\ - \frac{\omega_{pi}^2}{2k^2T_i/m_i}Z'\left(\frac{\omega/k}{\sqrt{2T_i/m_i}}\right) = 0.    
\end{multline}

Here, $\omega_{pe}$ and $\omega_{pi}$ are the electron and ion plasma frequencies respectively and $Z'$ is the derivative of the plasma dispersion function. We solve this relation for the parameters extracted from the Thomson scattering plotted in Fig.~\ref{Fig:TS}. At all times we find that the imaginary part of $\omega$ is negative and only becomes marginally positive for very small wavenumbers where the growth rate is too small to become important during the timescale of the experiment.

\textit{Appendix B: VPIC simulations of reconnection and LHDI} -- Two types of simulations were performed using VPIC: one of the global reconnection geometry and another of the generation of LHDW from the LHDI. Both simulations were performed in 2D, the former based on the simulations that were performed by Chien \textit{et al.} \cite{Chien_NatPhys_2023}, and the latter based on the work of Ng \textit{et al.} \cite{Ng_PRR_2024}. 

For the reconnection simulation a slice of the coils was represented in 2D Cartesian coordinates by two circular regions of current with a radius of 25 $\mu$m separated by $600\; \mu$m. Current was injected in this region, increasing linearly over a duration $t_R = 0.82\,\omega_{ci}^{-1}$ followed by an exponential decay with a characteristic $1/e$ time of $2\,\omega_{ci}^{-1}$ where $\omega_{ci}$ is the ion cyclotron frequency. Ions with a charge $Z = 18$ and mass 63.55 times the mass of a proton were injected alongside electrons from the boundary of the coil to represent the ionization of the coils. This was done with a linearly increasing density over $t_R$ with a normalizing density $n_0 = 10^{18}$ cm$^{-3}$, with no injection after $t_R$. A simulation with conditions matching those measured in the experiment is prohibitively expensive, therefore the simulation was scaled. An unphysically strong magnetic field was used to reduce the ratio of $\omega_{pe}/\omega_{ce}$ by a factor of 4 from the realistic value of 6.33 measured $150\;\mu$m from the coil toward the current sheet. The plasma $\beta = 0.063$ was maintained by increasing the electron and ion temperatures $T_e=T_i=6.4$ keV. The simulation was initialized with a uniform density of electrons and ions each with a density of $0.001n_0$ at 200 particles-per-cell. The box was $2.4\times2.4$ mm$^2$ with 2000 cells in each direction and absorbing boundaries for the particles and conducting boundaries for the fields. This corresponds to a resolution $dx=dz = 2L_D = d_e/4.4 = 1.27\rho_e$, where $L_D$ is the Debye length, and $\rho_e$ is the electron Larmor radius at $150\;\mu$m from the coil. 

The LHDI simulation was also performed in 2D Cartesian coordinates using the Vlasov confinement equilibrium outlined by Ng \textit{et al.} \cite{Ng_PRR_2024}. The simulation is set up with an ion density,
\begin{equation}
    n_i(x) = \frac{1 + \epsilon - \tanh{\left(\alpha \left(\frac{x^2}{a^2}-1 \right) \right)}}{1 + \epsilon + \tanh{(\alpha)}},
\end{equation}
\begin{equation}
    \epsilon = \frac{n_{\infty}(1 + \tanh(\alpha))}{1 - n_{\infty}}.
\end{equation}
Here $\alpha$ and $a$ are parameters used to set the thickness of the current layer and $n_{\infty}$ is used to set the asymptotic density. The density is normalized to $n_0$, the value on the left side of the box. The ions are initialized with a uniform temperature, while the electron temperature components $T_{xx}$ and $T_{zz}$ along with the electric and magnetic field are calculated recursively. While this simulation is used to study the dynamics of the LHDI, similar to the reconnection simulation, it cannot be performed using the realistic plasma conditions. An ion with a charge $Z = 1$ and a reduced mass equal to half that of a proton is used. The box had a size of $50d_e\times50d_e$ with $800\times900$ cells with reflecting boundary conditions for the particles and conducting boundaries for the fields on the lower and upper $x$ boundaries. Periodic boundary conditions were used for the upper and lower $y$ boundaries. The electrons and ions were both initialized with 1600 particles per cell and an ion to electron temperature ratio of 10. The electron plasma to cyclotron frequency $\omega_{pe}/\omega_{ce} = 22.1147$, the plasma beta on the left side of the current sheet $\beta = 7.65$, $\alpha = 1.3$, and $a = 25d_e$. Note, that while this plasma beta is relatively high, it is the local beta in the measurement region which is much larger than that in the reconnection upstream. The asymptotic density $n_\infty = 0.52n_0$ was chosen to set the electron drift velocity to thermal velocity ratio to the realistic value, $v_{e,y} = 0.169v_{th,e}$.

\end{document}